\documentstyle[prl,aps]{revtex}
\begin{document}
\draft

\title{Conductance fluctuations \\
at the fractional quantum Hall plateau transitions}

\author{Hae-Young Kee$^{a,b}$, Yong Baek Kim$^{a,b}$, 
Elihu Abrahams$^a$, and R. N. Bhatt$^{b,c}$}
\address{$^a$Serin Physics Laboratory, Rutgers University, 
Piscataway, NJ 08855-0849\\
$^b$Bell Laboratories, Lucent Technologies, Murray Hill, NJ 07974\\
$^c$Department of Electrical Engineering, Princeton University,
Princeton, NJ 08544}

\date{November 14, 1997}

\maketitle

\begin{abstract}

We obtain a ``mean field'' scaling flow of the longitudinal 
and the Hall conductivities in the fractional quantum Hall regime. 
Using the composite fermion picture and assuming that the composite 
fermions follow the Khmelnitskii-Pruisken scaling flow for the
integer quantum Hall effect, the unstable fixed points which govern 
the transitions between different fractional quantum Hall states 
are identified. Distributions of the critical longitudinal and
Hall conductivities at the unstable fixed points are obtained
and implications of the results for the experiments on mesoscopic 
quantum Hall systems are discussed. 
\end{abstract}


Many interesting phenomena in disordered mesoscopic metals
occur when the phase coherence length of the electrons exceeds the
sample size\cite{meso}. In particular, these mesoscopic metals show sample 
specific conductance fluctuations in contrast to macroscopic
systems where self-averaging leads to a disorder averaged
conductance\cite{meso,ucf}. 
It turns out that the magnitude of the fluctuation
is of the order of $e^2/h$ and universal in the sense that
it only depends on the symmetry of the problem. Thus, it has acquired
the name ``universal conductance fluctuations''\cite{ucf}.

On the other hand, in two-dimensional macroscopic electronic systems in
high perpendicular magnetic fields (quantum Hall effect regime), 
metallic behavior can be observed only near quantum 
phase transitions, 
{\it i.e.}, the transitions between different quantum 
Hall plateaus\cite{qhe}.
At these quantum critical points, the disorder
averaged longitudinal ($\sigma_{xx}$) and Hall 
conductivities ($\sigma_{xy}$) are expected 
to be universal\cite{klz,chk}. 
These universal critical conductivities can be
observed when the sample size becomes larger than the phase 
coherence length. 

For the integer quantum Hall effect, Khmelnitskii and then 
Pruisken\cite{khm} suggested a two parameter 
scaling flow in terms of $\sigma_{xx}$ and $\sigma_{xy}$. 
It was noticed that a non-linear sigma model with 
a topological term can describe the quantum phase transitions 
between different plateaus and the topological term is 
responsible for the metallic behavior at the quantum critical 
points\cite{khm}. 
According to this scaling flow, there are two types of
fixed points. There are stable fixed points which correspond to 
the integer quantum Hall states with $(\sigma_{xx},\sigma_{xy})=
(0,n)$ (in this paper, all the conductivities
are written in units of $e^2/h$). 
There are also unstable fixed
points which govern the critical behavior at the transitions
between adjacent integer quantum Hall states.
The disorder-averaged critical conductivities 
$(\langle \sigma_{xx} \rangle, \langle \sigma_{xy} \rangle)$ 
at the $(0,n-1) \rightarrow (0,n)$ transition or
at the corresponding unstable fixed points were 
suggested to be $(1/2, n-1/2)$. Here $n$ is a positive integer.
There exist numerical calculations\cite{bhatt1,bhatt2}
of $\langle \sigma_{xx} \rangle$
and $\langle \sigma_{xy} \rangle$ at the transition
$(\sigma_{xx},\sigma_{xy})=(0,0) \rightarrow (0,1)$,
which obtain
$\langle \sigma_{xx} \rangle \approx 0.5$
and $\langle \sigma_{xy} \rangle \approx 0.5$.

Besides the average value, it is of interest to determine
the conductance fluctuations near the critical point, which 
can be observed in mesoscopic quantum Hall
samples where the phase coherence length becomes larger than
the sample dimensions. 
More generally, one is interested in the
probability distribution of the conductivities at these 
quantum critical points, which may be expected to be universal.

A study of the Hall conductivity at the 
transition between the state with $(\sigma_{xx},\sigma_{xy})
=(0,0)$ and the integer quantum Hall state with 
$(\sigma_{xx},\sigma_{xy})=(0,1)$ was carried
out by Huo and Bhatt\cite{bhatt3}.
They found that the Hall conductivity distribution
is universal, independent of sample size, and symmetric
about $\sigma_{xy} \approx 0.5$, but has long power-law tails. 

Recently, experiments measuring two-terminal conductance provide
further motivation\cite{cobden}. 
Cobden and Kogan\cite{cobden} measured the 
two-terminal conductance (which corresponds to 
$\sigma_{xx}$\cite{comment})
of mesoscopic samples in the integer 
quantum Hall regime. They found large conductance 
fluctuations near the integer quantum Hall plateau
transitions indicating a broad distribution of the 
longitudinal conductance. In particular, they found that
the distribution is almost uniform in the interval between 
zero and one in units of $e^2/h$. 

There have been theoretical efforts to understand these
large fluctuations of $\sigma_{xx}$\cite{comment}. 
Wang, Jovanovic, and 
Lee\cite{wang} have calculated the ensemble averaged two-terminal
conductance and its fluctuations at the critical point
of the integer quantum Hall plateau transitions.
They used the Chalker-Coddington network model\cite{chalker} 
and periodic boundary conditions in the transverse direction.
They concluded that the average and all the higher
moments of the conductance distribution are
universal at the transition. It means that the
entire distribution is universal. At the same time,
the distribution turns out to be very broad in the 
sense that there is no well-defined typical value.
Cho and Fisher\cite{cho} also calculated
the distribution of the conductance in terms of 
the network model with periodic and open 
boundary conditions. It was explicitly shown that 
the conductance is more or less  uniformly distributed
between zero and one and there
is almost no weight for the conductances larger than one.
Thus, both of the results\cite{wang,cho} are consistent with the 
experiment of Cobden and Kogan\cite{cobden}.

These results naturally leads us to ask 
what the distributions of the longitudinal and the Hall 
conductivities are at the critical points for the {\it fractional} 
quantum Hall plateau transitions.
At best, the theoretical calculations mentioned above can be 
applied only to the integer quantum Hall effect because the 
electron-electron interaction is not included.
Since the electron-electron interaction is essential for
the fractional quantum Hall effect, the calculation of
the distributions of the conductivities at the critical
points in the fractional quantum Hall regime requires the
consideration of both the electron-electron interaction and 
disorder and is thus much more complicated.

Several years ago, Jain\cite{jain} proposed the composite 
fermion theory of the fractional quantum Hall effect. 
A composite fermion is obtained by attaching an even number 
$2m$ of fictitious flux quanta to an electron.
At the mean field level, one takes into account only 
the average of the fictitious magnetic field due to the 
attached fictitious magnetic flux.
Then the system can be described as fermions in an 
effective magnetic field $\Delta B = B - {\tilde B}$, 
where ${\tilde B} = 2m  n_e h c / e$
is the averaged fictitious magnetic field and $n_e$ is the
density of electrons. 
Therefore, in the mean field approximation, 
the fractional quantum Hall
states with $\nu = p / (2m p + 1)$ can be described 
as integer quantum Hall states of composite fermions with 
$p$ filled Landau levels occupied
in an effective magnetic field $\Delta B$\cite{jain,hlr}.
Here $\nu$ is the filling fraction and $p$ is an integer.
The most important correlation effects due to the
electron-electron interaction are supposed to be included in the 
construction of the composite fermions through the fictitious
flux quanta. 

In this paper, we use the composite fermion theory to get the
scaling flow of the longitudinal and Hall conductivities,
and the distributions of the conductivities at the critical 
points in the fractional quantum Hall regime. 
The main difference between the integer quantum Hall effect of
the electrons and that of the composite fermions is that 
the composite fermions experience both potential disorder 
and random flux disorder while the electrons have only
potential disorder\cite{hlr}. The random flux disorder for the composite
fermions arises due to the fact that the attached flux
moves together with the electron so that an inhomogeous electron
density distribution, which would occur in a random potential,
induces a random fictitious magnetic field.   
As a first step, in the absence of a rigorous study about the 
effects of both types of disorder, we assume that the integer 
quantum Hall effect of composite fermions follows the 
Khmelnitskii-Pruisken scaling flow\cite{khm}. 
Using a relation between 
the conductivity tensor of the electrons and that of the composite
fermions, we shall obtain the scaling flow in the fractional quantum Hall 
regime. In this scaling flow, the stable and the unstable 
fixed points are identified.  
Due to the mean-field nature of the connection between the integer
and fractional quantum Hall states mentioned above, we call it
a ``mean field" scaling flow of the fractional quantum Hall effect.
Assuming that the distributions of the critical conductivities 
of the composite fermions in their integer regime follow those of the 
electrons in their integer regime, we also get the distributions of 
the conductivities for the electrons at the
critical points of the fractional quantum Hall plateau transitions.

The relation between the resistivity tensor $\rho$ of the electrons
and that $\rho_{\rm cf}$ of the composite fermions is given by\cite{hlr}
\begin{equation}
\rho  = \rho^{\rm cf} + \rho^{\rm cs} \ ,
\end{equation}    
where
\begin{equation}
\rho^{\rm cs} = \pmatrix{ 0 & 2m \cr
-2m & 0} \ 
\end{equation}
comes from the Chern-Simons transformation.
Then  the longitudinal $\sigma_{xx}$ and the Hall $\sigma_{xy}$
conductivities of the electrons can be expressed in terms of the
composite fermion conductivities $\sigma^{\rm cf}_{xx}$ and 
$\sigma^{\rm cf}_{xy}$ as follows.   
\begin{eqnarray}
\sigma_{xx} &=& {\sigma^{\rm cf}_{xx} \ [ \ (\sigma^{\rm cf}_{xx})^2 +
(\sigma^{\rm cf}_{xy})^2 \ ] \over (\sigma^{\rm cf}_{xx})^2 +
[ \ \sigma^{\rm cf}_{xy} + 2m 
((\sigma^{\rm cf}_{xx})^2 + (\sigma^{\rm cf}_{xy})^2) \ ]^2} \cr
\sigma_{xy} &=& { \ [\sigma^{\rm cf}_{xy} + 
2m ((\sigma^{\rm cf}_{xx})^2 +
(\sigma^{\rm cf}_{xy})^2) \ ] \ [ \ (\sigma^{\rm cf}_{xx})^2 +
(\sigma^{\rm cf}_{xy})^2 \ ] \over (\sigma^{\rm cf}_{xx})^2 +
[ \ \sigma^{\rm cf}_{xy} + 2m 
((\sigma^{\rm cf}_{xx})^2 + (\sigma^{\rm cf}_{xy})^2) \ ]^2} \ .
\label{cond}
\end{eqnarray}
These equations are valid for any realization of the disorder.
For macroscopic samples, all the conductivities can be replaced
by the disorder averaged value if the self-averaging is legitimate. 

We assume that the integer quantum Hall effect of the 
composite fermions follows the Khmelnitskii-Pruisken scaling
flow in Fig.1\cite{khm}. 
There are stable fixed points 
$(\sigma^{\rm cf}_{xx},\sigma^{\rm cf}_{xy})=(0,n-1)$ which
represent the integer quantum Hall states of the composite 
fermions (here $n$ is a positive integer). 
There are also unstable fixed points
$(\sigma^{\rm cf}_{xx},\sigma^{\rm cf}_{xy})=(1/2,n-1/2)$
which control the transitions bewteen the quantum Hall states
$(\sigma^{\rm cf}_{xx},\sigma^{\rm cf}_{xy})=(0,n-1)$ and 
$(\sigma^{\rm cf}_{xx},\sigma^{\rm cf}_{xy})=(0,n)$. 

In order to get the scaling flow for the fractional quantum
Hall effect of the electrons, we use Eq.\ref{cond} to
map the Khmelnitskii-Pruisken flow in 
the $\sigma^{\rm cf}_{xx}-\sigma^{\rm cf}_{xy}$ 
plane to the flow in the $\sigma_{xx}-\sigma_{xy}$ plane.
We consider only the principal sequence $\nu = p/(2p+1)$ of the
fractional quantum Hall states, so $m = 1$ is taken
from now on.
Let us first calculate various limits.
For any $\sigma^{\rm cf}_{xy}$, if $\sigma^{\rm cf}_{xx} \rightarrow
\infty$, then $\sigma_{xx} \rightarrow 0$ and 
$\sigma_{xy} \rightarrow 1/2$. Thus, the entire line 
$\sigma^{\rm cf}_{xx} \rightarrow \infty$ goes to the point
$(\sigma_{xx},\sigma_{xy})=(0,1/2)$.
The line $\sigma^{\rm cf}_{xx}=0$ 
goes to $(\sigma_{xx},\sigma_{xy})=(0,\sigma^{\rm cf}_{xy}/
(1 + 2\sigma^{\rm cf}_{xy}))$.
Thus all the stable fixed points
$(\sigma^{\rm cf}_{xx},\sigma^{\rm cf}_{xy})=(0,n-1)$ go to
\begin{equation}
(\sigma_{xx},\sigma_{xy})=\left ( 0, {n-1 \over 2(n-1)+1} \right )
\end{equation}
which become also the stable fixed points in the 
$\sigma_{xx}-\sigma_{xy}$ plane and correspond to the fractional 
quantum Hall states of the electrons. 
On the other hand, all the unstable fixed points
$(\sigma^{\rm cf}_{xx},\sigma^{\rm cf}_{xy})=(1/2,n-1/2)$
can be mapped to
\begin{equation}
(\sigma_{xx},\sigma_{xy})=\left ( {1 \over 2}{4n^2-4n+2 \over 
1 + (4n^2-2n+1)^2}, {1 \over 2}{(4n^2-4n+2)(4n^2-2n+1) 
\over 1+(4n^2-2n+1)^2} \right ) \ .
\label{ufix}
\end{equation}
For example, $(\sigma^{\rm cf}_{xx},\sigma^{\rm cf}_{xy})
=(1/2,1/2)$, $(1/2,3/2)$, and $(1/2,5/2)$
are mapped to $(\sigma_{xx},\sigma_{xy})=(1/10,3/10)$, 
$(1/34,13/34)$, and $(13/962,403/962)$.
These are the critical conductivities at the transitions
$(\sigma_{xx},\sigma_{xy})=(0,0) \rightarrow (0,1/3)$, 
$(0,1/3) \rightarrow (0,2/5)$, and $(0,2/5) \rightarrow (0,3/7)$
respectively.
These critical conductivities have been also calculated in
previous studies and are supposed to be valid for macroscopic
samples\cite{klz,chk}.
As a last exercise, it can be easily seen that
$(\sigma^{\rm cf}_{xx},\sigma^{\rm cf}_{xy})=(0,n-1/2)$
are mapped to 
\begin{equation}
(\sigma_{xx},\sigma_{xy})=\left ( 0,{2n-1 \over 4n} \right ) \ .
\end{equation}  

The resulting scaling flow for the principal sequence 
$\nu = p/(2p+1)$ is shown in Fig.2.
It looks quite similar to the case of the integer quantum
Hall states and is consistent with the selection rules in
the law of corresponding states proposed by Kivelson, 
Lee, and Zhang\cite{klz}. For example, the direct transition between
$(\sigma_{xx},\sigma_{xy})=(0,0)$ and $(0,2/5)$ states
is not allowed. The scaling flow also gives us some new
information. For example, the scaling curve which starts at 
$(\sigma_{xx},\sigma_{xy})=(0,1/2)$
and goes to $(1/10,3/10)$ has
a maximum at $(\sigma_{xx},\sigma_{xy})=(1/8,3/8)$.
This implies that the bare longitudinal conductivity of the
smaple should be smaller than $1/8$ in
order to see the $\nu=1/3$ quantum Hall state.
It is also interesting to notice that our result for the 
scaling flow in the fractional quantum Hall regime is 
different from that proposed by Laughlin {\it et al.} 
some years ago\cite{laughlin}.

Now we consider the statistical properties of the 
conductivities at the critical points which are governed
by the unstable fixed points given by Eq.\ref{ufix}.
Let us assume that the distributions of the critical 
conductivities for the integer quantum Hall effect of the
composite fermions is the same as those for the electrons.
Then the distribution of the critical longitudinal conductivities 
for the composite fermions at 
$(\sigma^{\rm cf}_{xx},\sigma^{\rm cf}_{xy})=(1/2,n-1/2)$
will be taken as\cite{cobden,comment,wang,cho}
\begin{eqnarray}
P(\sigma^{\rm cf}_{xx}) &=& \cases{ 1 & 
\hskip 1.0cm $0 \le \sigma^{\rm cf}_{xx} \le 1$ \cr 
0 & \hskip 1.0cm $\sigma^{\rm cf}_{xx} > 1$  \ . \cr} 
\end{eqnarray}
On the other hand, the distribution of the Hall conductivity 
for the composite fermions is taken as
\begin{eqnarray}
P(\sigma^{\rm cf}_{xy}) &=& \cases{ 1.246
e^{-5.309 (\sigma^{\rm cf}_{xy}-
\langle \sigma^{\rm cf}_{xy} \rangle )^2}
\left [ 1 - 0.2791 (\sigma^{\rm cf}_{xy}-
\langle \sigma^{\rm cf}_{xy} \rangle)^2 \right ] & 
$| \sigma^{\rm cf}_{xy} - 
\langle \sigma^{\rm cf}_{xy} \rangle | \le 0.5084$ \cr 
0.04122 / (\sigma^{\rm cf}_{xy}-
\langle \sigma^{\rm cf}_{xy} \rangle)^{2.9}
& $| \sigma^{\rm cf}_{xy} - 
\langle \sigma^{\rm cf}_{xy} \rangle | >  0.5084$ \ , \cr} 
\label{numerical}
\end{eqnarray}
where $\langle \sigma^{\rm cf}_{xy} \rangle = n-1/2$.
This distribution turns out to be an excellent parametrization
of the numerical result\cite{bhatt3}.
Furthermore, both the distribution and its derivative at 
$| \sigma^{\rm cf}_{xy} - \langle \sigma^{\rm cf}_{xy} 
\rangle | =  0.5084$ are continuous.
Notice that the same distribution is taken for any $n$ with
$\langle \sigma^{\rm cf}_{xy} \rangle = n - 1/2$ 
due to the invariance of the effective
action or the non-linear sigma model\cite{khm} under
$\sigma^{\rm cf}_{xy} \rightarrow \sigma^{\rm cf}_{xy} + l$,
where $l$ is an integer.
One can also see that $P(\sigma^{\rm cf}_{xy})$ does not
have a second moment due to the long tail.

Notice that the distributions of 
$\sigma^{\rm cf}_{xx}$ and $\sigma^{\rm cf}_{xy}$ may be 
correlated. In the absence of a rigorous study about the 
relation between these two distributions, in this paper,
we assume that they are independent each other for simplicity.
Then the distribution of $\sigma_{xx}$ and $\sigma_{xy}$ at the
critical points of the fractional quantum Hall plateau
transitions can be obtained from the convolution of 
$P(\sigma^{\rm cf}_{xx})$ and $P(\sigma^{\rm cf}_{xy})$
using Eq.\ref{cond}.
That is, the distribution $P(\sigma_{xx})$ and $P(\sigma_{xy})$
of $\sigma_{xx}$ and $\sigma_{xy}$ are given by
\begin{eqnarray}
P(\sigma_{xx}) &=& \int^{\infty}_{-\infty} d \sigma_{xy}
|J(\sigma_{xx},\sigma_{xy}:\sigma^{\rm cf}_{xx},\sigma^{\rm cf}_{xy})|
P(\sigma^{\rm cf}_{xx})P(\sigma^{\rm cf}_{xy}) \cr
P(\sigma_{xy}) &=& \int^{\infty}_{-\infty} d \sigma_{xx}
|J(\sigma_{xx},\sigma_{xy}:\sigma^{\rm cf}_{xx},\sigma^{\rm cf}_{xy})|
P(\sigma^{\rm cf}_{xx})P(\sigma^{\rm cf}_{xy}) \ ,
\end{eqnarray}
where $\sigma^{\rm cf}_{xx}$ and $\sigma^{\rm cf}_{xy}$ in the
integrand should be written in terms of $\sigma_{xx}$ and 
$\sigma_{xy}$ from Eq.\ref{cond}.
Here $J(\sigma_{xx},\sigma_{xy}:\sigma^{\rm cf}_{xx},
\sigma^{\rm cf}_{xy})$ is the Jacobian for the change of
the variables from $\sigma^{\rm cf}_{xx}$ and $\sigma^{\rm cf}_{xy}$
to $\sigma_{xx}$ and $\sigma_{xy}$.

The results $P(\sigma_{xx})$ and $P(\sigma_{xy})$ at the critical
point (or the unstable fixed point) for the transition 
$(\sigma_{xx},\sigma_{xy})=(0,0) \rightarrow (0,1/3)$ 
are shown in Fig.3 (a) and (b).
One can see that $P(\sigma_{xx})$ has a maximum at $\sigma^{\rm max}_{xx}
\approx 0.09$, but the distribution is still broad so that this is
not really a typical value. Notice that there is almost no weight 
beyond $\sigma_{xx} \approx 0.5$.
The average and the second moment are given by 
$\langle \sigma_{xx} \rangle = 0.1$ and 
$\delta \sigma_{xx} = \sqrt{\langle \sigma^2_{xx} \rangle - 
\langle \sigma_{xx} \rangle^2} = 0.0725$. Thus the relative fluctuation is
$\delta \sigma_{xx} / \langle \sigma_{xx} \rangle = 0.725$.
On the other hand, $P(\sigma_{xy})$ has a well-defined typical
value $\sigma^{\rm typical}_{xy} \approx 0.37$ which is different
from the average value $\langle \sigma_{xy} \rangle = 0.3$. 
The reason why $P(\sigma_{xy})$ is skewed is that the relation 
Eq.\ref{cond} between $\sigma_{xx}, \sigma_{xy}$ and 
$\sigma^{\rm cf}_{xx}, \sigma^{\rm cf}_{xy}$ is non-linear.
The cusp-like feature near $\sigma^{\rm typical}_{xy}$ is not
a cusp and we can show that it is perfectly smooth if one plots
$P(\sigma_{xy})$ for a limited interval, {\it e.g.}, between
0.35 and 0.4.
The second moment and the relative fluctuation for $\sigma_{xy}$
are given by $\delta \sigma_{xy} = 0.0675$ and 
$\delta \sigma_{xy} / \langle \sigma_{xy} \rangle = 0.225$
respectively.

For the transition 
$(\sigma_{xx},\sigma_{xy})=(0,1/3) \rightarrow (0,2/5)$,
the results are shown in Fig.3 (c) and (d).
$P(\sigma_{xx})$ is quite broad up to $0.1$ with a maximum at
$\sigma^{\rm max}_{xx} \approx 0.033$ , but 
$P(\sigma_{xy})$ has a well-defined typical value 
$\sigma^{\rm typical}_{xy} \approx 0.39$
(note that $\langle \sigma_{xy} \rangle = 0.382$).    
The average and the second moment for $\sigma_{xx}$ are given by 
$\langle \sigma_{xx} \rangle = 0.0294$ and 
$\delta \sigma_{xx} = 0.0184$. 
Thus the relative fluctuation is
$\delta \sigma_{xx} / \langle \sigma_{xx} \rangle = 0.640$.
On the other hand, the second moment and the relative fluctuation for
$\sigma_{xy}$ are given by
$\delta \sigma_{xy} = 0.0220$ and 
$\delta \sigma_{xy} / \langle \sigma_{xy} \rangle = 0.0581$.

As the last example, we consider the transition
$(\sigma_{xx},\sigma_{xy})=(0,2/5) \rightarrow (0,3/7)$,
the results are shown in Fig.3 (e) and (f).
$P(\sigma_{xx})$ is again very broad up to $0.04$ with a 
maximum at $\sigma^{\rm max}_{xx} \approx 0.0164$, but 
$P(\sigma_{xy})$ has a well-defined typical value 
$\sigma^{\rm typical}_{xy} \approx 0.42$
while $\langle \sigma_{xy} \rangle = 0.419$.    
The average and the second moment for $\sigma_{xx}$ are given by 
$\langle \sigma_{xx} \rangle = 0.0135$ and 
$\delta \sigma_{xx} = 0.00817$. Thus the relative fluctuation is
$\delta \sigma_{xx} / \langle \sigma_{xx} \rangle = 0.613$.
For $\sigma_{xy}$, the second moment and the relative fluctuation
are given by
$\delta \sigma_{xy} = 0.0111$ and 
$\delta \sigma_{xy} / \langle \sigma_{xy} \rangle = 0.0266$.
All these results are summarized in Table 1.

Notice that both the relative fluctuations of $\sigma_{xx}$
and of $\sigma_{xy}$ are decreasing as $n$ increases
at the critical points given by Eq.\ref{ufix}.
However, the change of magnitude in the case of 
$\sigma_{xy}$ is quite large while that in the case of 
$\sigma_{xx}$ is very small.
This implies that the distributions of $\sigma_{xx}$ are
almost equally broad for any $n$, but those of $\sigma_{xy}$ 
become sharper and sharper as $n$ is increased.
Therefore, in experiments, it is expected that there will be large 
conductance fluctuations in $\sigma_{xx}$ at any fractional
quantum Hall plateau transition while the conductance
fluctuations in $\sigma_{xy}$ become smaller and smaller 
as the filling fraction of the quantum Hall states involved
in the transition approaches ${1 \over 2m} {e^2 \over h}$,
where $m$ is a positive integer.

We also calculated $P(\sigma_{xx})$ and $P(\sigma_{xy})$ using
a Gaussian distribution for $\sigma^{\rm cf}_{xy}$.
The purpose was to examine the effects of the long tail in
the distribution obtained from the numerical calculation
which we used above (see Eq.\ref{numerical}).
Comparing the results in the cases of using the Gaussian and the 
numerically-calculated distribution for $\sigma^{\rm cf}_{xy}$, 
we found that, in the latter case, the tails of $P(\sigma_{xx})$ and 
$P(\sigma_{xy})$ become indeed longer than those in the Gaussian case.
However, the contributions from the tails are not large
enough to give a diverging second moment so that the second
moment of $P(\sigma_{xy})$ exists at the fractional quantum Hall 
plateau transitions in contrast to the case of the distribution 
of the Hall conductivity at the integer quantum Hall plateau 
transitions\cite{bhatt3}.
Other than these slightly longer tails, there is no 
qualitative difference between the two cases in the resulting
$P(\sigma_{xx})$ and $P(\sigma_{xy})$. 

In summary, a ``mean field'' scaling flow for the longitudinal
and the Hall conductivities is obtained from the composite
fermion theory assuming that the integer quantum Hall effect
of the composite fermions follows the Khmelnitskii-Pruisken 
scaling flow. At the unstable fixed points which govern the
transitions between the different fractional quantum Hall 
states, the distributions of the critical conductivities 
are obtained. It is found that the distributions of the
longitudinal conductivity still remain as being
broad and having almost no weight beyond certain value of the
conductivity. The distributions of the Hall conductivity
have well-defined typical values which are generally 
different from the average values and become sharper and
sharper as the average Hall conductivity at the 
critical points approaches ${1 \over 2m} {e^2 \over h}$.

Y.B.K. thanks P. A. Lee for motivating the present problem 
and D. E. Khmelnitskii for helpful discussion. 
This work was supported by NSF grant No. DMR-96-32294 
(H.Y.K., Y.B.K., and E.A.) and DMR-9400362 (R.N.B.).
H.Y.K. was also supported by the Korea Research Foundation.
R.N.B. thanks the J. S. Guggenheim foundation for a
fellowship and Bell Laboratories for hospitality at the 
early stage of this project.

\begin{table}
\caption{Summary of the results for the distributions of
$\sigma_{xx}$ and $\sigma_{xy}$ at the fractional quantum 
Hall plateau transitions}
\begin{tabular}{|l|l|l|l|l|l|l|}
{} & $\langle \sigma_{xx} \rangle$ & $\langle \sigma_{xy} \rangle$ &
$\delta \sigma_{xx}$ & $\delta \sigma_{xy}$ &
$\delta \sigma_{xx} / \langle \sigma_{xx} \rangle$ & 
$\delta \sigma_{xy} / \langle \sigma_{xy} \rangle$ \\ 
\hline \hline
$0 \rightarrow 1/3$ & 
0.1 & 0.3 & 0.0725 & 0.0675 & 0.725 & 0.225 \\ 
\hline
$1/3 \rightarrow 2/5$ & 
0.0294 & 0.382 & 0.0184 & 0.0220 & 0.640 & 0.0581 \\ 
\hline
$2/5 \rightarrow 3/7$ &
0.0135 & 0.419 & 0.00817 & 0.0111 & 0.613 & 0.0266 \\ 
\end{tabular}
\end{table} 

\begin{figure}
\caption{The Khmelnitskii-Pruisken scaling flow for the 
integer quantum Hall effect. In this paper, it is assumed 
that it can be applied to the integer quantum Hall effect 
of the composite fermions. $\Box$ and $\bigcirc$ correspond
to the unstable and the stable fixed points.
}
\end{figure}

\begin{figure}
\caption{A ``mean field'' scaling flow for the fractional 
quantum Hall effect of the electrons. 
$\Box$ and $\bigcirc$ correspond
to the unstable and the stable fixed points.
We show only the part which has $\sigma_{xy} \le 1/2$. 
}
\end{figure}

\begin{figure}
\caption{(a), (c), and (e) are $P(\sigma_{xx})$s 
while (b), (d), and (f) are $P(\sigma_{xy})$s for the
transitions $(\sigma_{xx},\sigma_{xy})=(0,0) \rightarrow
(0,1/3)$, $(0,1/3) \rightarrow (0,2/5)$, and 
$(0,2/5) \rightarrow (0,3/7)$ respectively.
}
\end{figure}
  

\begin{references}

\bibitem{meso} {\it Mesoscopic Phenomena in Solids}, edited by
B. L. Altshuler, P. A. Lee, and R. A. Webb (North-Holland, 
New York, 1991).
\bibitem{ucf} P. A. Lee, A. D. Stone, and H. Fukuyama, Phys. 
Rev. B {\bf 35}, 1039 (1987), and references therein.
\bibitem{qhe} For reviews, see B. Huckenstein, Rev. Mod. Phys.
{\bf 67} 357 (1995); {\it The Quantum Hall Effect}, edited by
R. E. Prange and S. M. Girvin (Springer-Verlag, New York, 1990).
\bibitem{klz} S. A. Kivelson, D.-H. Lee, and S.-C. Zhang, 
Phys. Rev. B {\bf 46}, 2223 (1992).
\bibitem{chk} D. B. Chklovskii and P. A. Lee, 
Phys. Rev. B {\bf 48}, 18060 (1993).
\bibitem{khm} D. E. Khmelnitskii, JETP Lett {\bf 38}, 552 (1983);
A. M. M. Pruisken, Nucl. Phys. {\bf B 235}, 277 (1984).
\bibitem{bhatt1} Y. Huo, R. E. Hetzel, and R. N. Bhatt, 
Phys. Rev. Lett. {\bf 70}, 481 (1993).
\bibitem{bhatt2} Y. Huo and R. N. Bhatt, 
Phys. Rev. Lett. {\bf 68}, 1375 (1992).
\bibitem{bhatt3} Y. Huo, Ph.D. Thesis 
(Princeton University, January 1994, unpublished), 
Y. Huo and R. N. Bhatt (Preprint, 1997).
\bibitem{cobden} D. H. Cobden and E. Kogan, Report No. 
cond-mat/9606114, 1996 (to be published).
\bibitem{comment} In general, two-terminal conductance is not 
the same as the longitudinal conductance obtained from the
Kubo formula or four-terminal measurement. 
Here we assume that the major contribution to
the two-terminal conductance comes from the logitudinal 
conductance. Thus we take the distribution of the two-terminal
conductance as that of the logitudinal conductance. 
Even if the two-terminal conductance is quite different from 
the longitudinal conductance, we expect that we will get 
qualitatively the same results. 
\bibitem{wang} Z. Wang, B. Jovanovic, and D.-H. Lee, 
Phys. Rev. Lett. {\bf 77}, 4426 (1996).
\bibitem{chalker} J. T. Chalker and P. D. Coddington, 
J. Phys. C {\bf 21}, 2665 (1988).
\bibitem{cho} S. Cho and M. P. A. Fisher, 
Phys. Rev. B {\bf 55}, 1637 (1997).
\bibitem{jain} J. K. Jain, Phys. Rev. Lett. {\bf 63}, 
199 (1989); A. Lopez and E. H. Fradkin, 
Phys. Rev. B {\bf 44}, 5246 (1991). 
\bibitem{hlr} B. I. Halperin, P. A. Lee and N. Read, 
Phys. Rev. B {\bf 47}, 7312 (1993).
\bibitem{laughlin} R. B. Laughlin {\it et al.}, 
Phys. Rev. B {\bf 32}, 1311 (1985).

\end{references}
\end{document}